\documentclass[aps,prl,twocolumn,showpacs,groupedaddress]{revtex4}
\usepackage{amsmath}
\usepackage{graphicx}
\usepackage{dcolumn}
\usepackage{bm}
\usepackage{color}

\begin{document}
\title{Comment on ``Molybdenum at High Pressure and Temperature: Melting
from Another Solid Phase''}
\author{C. Cazorla$^{{\rm a},{\rm b},{\rm c}}$~, D. Alf\`e$^{{\rm a},{\rm b},{\rm c},{\rm d}}$~,
and M. J. Gillan$^{{\rm a},{\rm b},{\rm c}}$}
\affiliation{
$^{\rm a}$ London Centre for Nanotechnology, UCL, London WC1H OAH, UK \\
$^{\rm b}$ Department of Physics and Astronomy, UCL, London WC1E 6BT, UK\\
$^{\rm c}$ Materials Simulation Laboratory, London WC1E 6BT, U.K. \\
$^{\rm d}$ Department of Earth Sciences, UCL, London, WC1E 6BT, U.K.}
\maketitle
There has been a major controversy over the past seven years about the high-pressure
melting curves of transition metals. Static compression (diamond-anvil cell:~DAC) experiments up to the Mbar 
region give very low melting slopes $dT_{m}/dP$~, but shock-wave (SW) data reveal transitions
indicating much larger $dT_{m}/dP$ values. \emph{Ab initio} calculations support the 
correctness of the shock data. In a very recent Letter~[\onlinecite{belonoshko08}]~, 
Belonoshko \emph{et al.} propose a simple and elegant resolution of this conflict  
for molybdenum. Using \emph{ab initio} calculations based on density functional theory (DFT)~,
they show that the high-$P$/high-$T$ phase diagram of Mo must be more complex than was hitherto
thought. Their calculations give convincing evidence that there is a transition boundary
between the normal bcc structure of Mo and a high-$T$ phase, which they suggest could be fcc.
They propose that this transition was misinterpreted as melting in DAC experiments~[\onlinecite{errandonea01}].
In confirmation, they note that their boundary also explains a transition seen in the SW data~[\onlinecite{hixson89}]. 
We regard Belonoshko \emph{et al.}'s Letter as extremely important, but we note that it raises
some puzzling questions, and we believe that their proposed phase diagram cannot be completely 
correct. Since the neighbors of Mo in the $4d$ series are the hcp-structured Tc and Ru, we would expect 
Mo to transform more readily to the hcp structure than to the fcc structure suggested
by Belonoshko \emph{et al.} Indeed, if their phase diagram were correct, it would imply that at
$P \ge 7$~Mbar, there are two separate stability fields of fcc, which would be highly unusual. 
Another puzzle is that in the DAC experiments~[\onlinecite{errandonea01}]~, the samples are reported to be in 
a liquid-like state on the high-$T$ side of the boundary where Belonoshko \emph{et al.} propose
that the material is crystalline. 

To check whether hcp, rather than fcc, could be the stable 
phase at high-$P$/high-$T$~, we have calculated the Helmholtz and Gibbs free energies of the 
bcc, fcc and hcp phases of Mo, using essentially the same quasiharmonic methods as used 
by Belonoshko \emph{et al.} The phonon frequencies in our calculations were obtained using 
the small-displacement method~[\onlinecite{kresse95}]~, with large supercells of 216 atoms and full inclusion of 
thermal electronic excitations. Our calculated bcc-fcc boundary for $P > 3.5$~Mbar, where 
fcc is vibrationally stable, agrees very closely with that of Belonoshko \emph{et al.}~ However, 
we find that hcp is noticeably more stable in this region and our bcc-hcp boundary is reported in Fig.~1~.
Surprisingly, the bcc-fcc slope $dT/dP$ is higher than the bcc-hcp one, so that
there must be a bcc-fcc-hcp triple point at $P \simeq 2.9$~Mbar and $T \simeq 4800$~K. The
resulting schematic phase diagram of Mo shown in Fig.~1 indicates that the transition seen in 
DAC experiments is indeed bcc-fcc. Concerning the possible liquid-like behaviour of the crystal
above the bcc-fcc boundary, we comment on the harmonic vibrational instability of
fcc below $3.5$~Mbar found by Belonoshko \emph{et al.} Our calculations confirm this instability,
and show that it is associated with a negative shear elastic constant. We find that hcp is also
elastically unstable at $T = 0$~K and $P < 3.5$~Mbar. Above the bcc-fcc boundary, the crystal
must therefore be highly anharmonic and stabilized by entropy. The implication is that above
this boundary the fcc and hcp phases have almost exactly the same free energy and both are on the
verge of elastic instability. It seems highly likely that these unusual features underly the 
controversial interpretation of the static compression measurements.  
\begin{figure}[t]
\vspace{-0.7cm}
\centerline{
        \includegraphics[width=0.90\linewidth,angle=0]{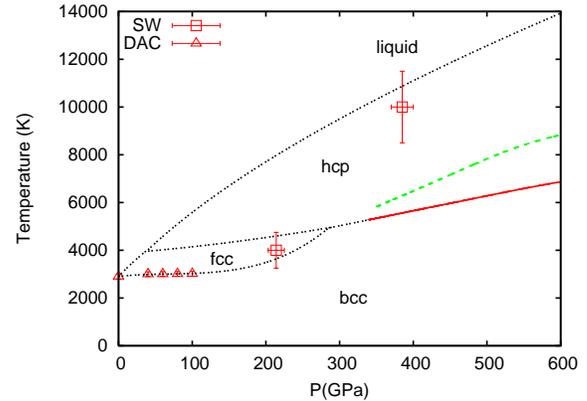}}%
        \vspace{-0.35cm}
        \caption{Schematic phase diagram of Mo. Dashed (green) curve: bcc-fcc boundary from Ref.~[\onlinecite{belonoshko08}]~;
	Solid (red) curve: bcc-hcp boundary from present calculations; melting curve is adapted from Ref.~[\onlinecite{belonoshko08}];
	dotted boundaries are estimated.}
\vspace{-0.5cm}
\end{figure}

\end{document}